\documentclass[twocolumn,showpacs,preprintnumbers,amsmath,amssymb]{revtex4}

\usepackage{graphicx}
\usepackage{dcolumn}
\usepackage{bm}

\usepackage{epsfig}
\begin{document}

\title{\bf Numeric simulation of relativistic stellar core collapse and the formation of Reissner-Nordstrom black holes}

\author{Cristian R. Ghezzi}
\email{ghezzi@ime.unicamp.br}
\author{Patricio S. Letelier}
\email{letelier@ime.unicamp.br}

\address{\it Department of Applied Mathematics, 
Instituto de Matem\'atica, Estat\'{\i}stica e Computa\c c\~ao Cient\'{\i}fica,\\ 
Universidade Estadual de Campinas,\\ Campinas, S\~ao Paulo, Brazil}

\date{\today}

\begin{abstract}

The time evolution of a set of $22 M_\odot$  unstable charged 
stars that collapse  
is computed integrating the Einstein-Maxwell equations.
The model simulate
the collapse of an spherical star that had exhausted its nuclear fuel and have 
or acquires a
net  electric charge in its core while collapsing. 
When the 
charge to mass ratio is $Q/\sqrt{G}M \ge 1$ the star do not collapse and spreads.
On the other hand, it is observed a different physical 
behavior with a charge to mass ratio $1 > Q/\sqrt{G} M > 0.1$. In this case, 
the collapsing matter forms a bubble enclosing a lower density core. 
We discuss an immediate astrophysical consequence 
of these results that is a more efficient neutrino trapping during the
 stellar collapse and an alternative mechanism for powerful supernova explosions. 
The outer space-time of the star is the Reissner-Nordstr\"om solution that match smoothly
with our interior numerical solution, thus the collapsing models forms Reissner-Nordstr\"om black holes.

\end{abstract}
\pacs{04.25.Dm;04.40.Nr;04.40.Dg;04.70.Bw;95.30.Sf;97.60.Bw;97.60.Lf}

\maketitle
\section{INTRODUCTION}
In this work we intend to study the effects of an electric field on the collapse
 of a massive star. 
We perform direct relativistic simulations assuming spherical symmetry and integrating the Einstein-Maxwell equations.
We studied the collapse of several stars with different values of the total electric charge
and determined the limits at which the collapse is avoided by the effect of the repulsive electric field. The electric charge is carried by the 
particles that compose the collapsing fluid.
The electromagnetic field and the internal energy of the gas 
contributes to the total mass-energy of the star, 
so it is not clear whether it is possible to overcharge a collapsing configuration.
 The analysis and conclusions drawn for collapsing charged shells  cannot
be directly extrapolated for this case,
and  a complete relativistic calculation performed in a self-consistent way is needed 
to know the outcome of the charged collapse.

In particular, we found the formation of a bubble that was not predicted before.
Although its theoretical explanation is quite natural, the formation of the bubble depends on 
the initial conditions and its evolution is far from obvious due to the non-linearities 
of the Einstein-Maxwell equations.
Numerical solutions for the problem of charged
stellar collapse were found only recently \cite{ghezzi2}, \footnote{ 
There are in the literature numerical and analytical solutions of 
 charged spherical stars in hydrostatic 
equilibrium (See for example Refs. \cite{ray}, \cite{defelice}, \cite{anninos}).
 There are also numerical research 
 on the collapse of charged scalar fields, with particular emphasis on the mass inflation effect
near the Cauchy horizon  \cite{piran} (and references therein), 
and global analysis on the mass inflation effect on the stellar
collapse \cite{poisson}.
The physics of charged thin shells collapse had been analyzed in \cite{israel}, and \cite{boulware}.
The equations for stellar charged  collapse had been obtained in the references \cite{bekenstein}, \cite{mashhoon},
and specially for their numeric integration in \cite{ghezzi2}.}.

It is commonly assumed  that
the stars are nearly neutrally charged due to a mechanism of selective accretion
of charges from the surrounding interstellar medium.
Therefore, if a star is initially positively (negatively) charged, accretion
of negative (positive) charges from the surrounding gas will tend to neutralize
the total net charge.  
The same reasoning is applied to black holes \cite{wald}. 
We observe that  the effect of selective accretion were never fully calculated to our knowledge.
On the other hand, an star can be electrically neutral while 
possessing   a huge internal electric field \footnote{The total charge of a 
star can be zero while its internal  electric field is huge.
An example of this are charge separation effects in  compact stars, see Refs.
\cite{olson1}, \cite{olson2}; charge separation during accretion of matter onto a star
or onto a black hole,
see Ref. \cite{shvartsman} and \cite{diego}; 
electric field generation due to magneto-rotational effects \cite{punsly};
electric field in pulsars \cite{goldreich};
or net charge effects due to physics in extra-dimensions \cite{mosquera}.}. 

The internal electric field of an astronomical object can be very large in some special scenarios \cite{zeldovich}, \cite{olson1}, 
\cite{olson2}, \cite{shvartsman}.
In particular, the charge separation inside  a spherical compact star, in hydrostatic equilibrium, 
can be very large  when  one of the plasma components is a degenerate
gas while the other is a Maxwell-Boltzmann gas, i.e., 
like the gas of degenerated electrons and the gas of nuclei in a white dwarf \cite{olson2}.

The  calculations presented in this paper corresponds to the case in which the star posses a net electric charge.
However the results can be taken as  an approximation to the more complicated problem of an 
 star with total zero charge and with a non zero  internal electric field.
We leave this point aside and for clarity  we will concentrate here on the dynamics of an star with a total positive charge.

We ask whether an electrically charged  star can collapse to form a charged black hole.
Would the Coulomb repulsion avoid the collapse of the star ?, and moreover, are there  
physical differences with the physics of uncharged collapse ?.  
Moreover, from pure analytic analysis is not clear if it is possible to form an overcharged black hole (with $Q >\sqrt{G}M$).
The dynamics of a collapsing star could be quite different from the collapse of a  charged shell
onto an already formed Reissner-Nordstr\"om black hole, mainly due to backreaction effects.
In the present
paper, we will try to give the answer to these questions.

Extremely charged black holes ($Q/\sqrt{G}M = 1$) represent an extreme limit in the context of
the cosmic censorship hypothesis, since bodies
with charge equal or higher  than extremal are undressed by event horizons and constitutes naked singularities
 \cite{joshi}, \cite{wald}.
On the other hand, the formation of Reissner-Nordstr\"om (RN) black holes, is in connection
with the third law of black hole thermodynamics \cite{anninos}, \cite{israel}, \cite{ghezzi2}, because
an extreme RN black hole has zero temperature.
So, there is an interest on the formation of charged black holes from a pure
theoretical point of view \footnote{
Electromagnetic fields  plays an important role in several models for the inner engine of  
gamma ray bursts. During the 
accretion of plasma onto a black hole the electromagnetic field can annihilate
above a certain limit
due to quantum effects, and
produce a ``dyadosphere'' of electron-positron pairs  that in turn annihilates   to produce
a  gamma ray burst  \cite{ruffini}.}. 

In this paper we are not concerned with the mechanism that produce charge or an internal electric field.
Neither magnetic fields nor
rotation are taken into account, although they could increase the magnitude of the electric fields.

Without loss of generality, the density of charge was chosen proportional to the rest mass density. 
The interior solutions found can be matched smoothly with the
 Reissner-Nordstr\"om exact solution for the vacuum space-time \cite{ghezzi2}. 
Therefore, in the cases in which the star collapses the result is 
 the formation of a Reissner-Nordstr\"om black hole. 

The paper is organized as follows. In Section II, we discuss the problem of charged 
collapse analyzing the  order of magnitude of the physical quantities. In the Section III
 are described the general relativistic equations
that govern the dynamics of the stellar core collapse. In Section IV,  are described the numerical techniques
used to integrate the general relativistic equations and some caveats on its applications. 
Section V brings a discussion and interpretation of the numerical results. In the Subsection A, we show the 
calculation of the maximum mass formula for Newtonian charged stars, and in Subsection B we calculate 
the optical depth for the neutrinos emitted during the collapse.
We also discuss the implications for core collapse supernova. We end in Section VI presenting some final
 remarks.

\section{Some order of magnitude estimates}
In this section we show some order of magnitude estimates,  in order
to put in a clear perspective the problem of a self-gravitating charged fluid sphere.
The calculations in the present section are valid only in the non-relativistic regime.

The formation of a charged star,
is possible when the
gravitational attraction  overwhelms the electrostatic repulsion on each single particle of
the gas that is collapsing, i.e.,
\begin{eqnarray}
F_{grav} \ge F_{elect}\,,
\end{eqnarray}
or equivalently
\begin{eqnarray}
\label{cargas}
q\,Q_{s} \ge m\,M_{s}\,,
\end{eqnarray}
where $Q_{s}$ and $M_{s}$ are the charge and mass of the star and $q$ and $m$ are the charge 
and mass of the particles.
In this section we use a Newtonian approach, the full relativistic case could be different. 

Assume a star that has a total charge
to mass ratio $Q_{s}/\sqrt{G} M_{s}$. According to Eq. (\ref{cargas})
any particle with a charge to mass ratio $q/\sqrt{G} m$  can be added
to the star if 
\begin{equation}
\label{limit1}
\alpha_1 \equiv \frac{Q_{s}}{\sqrt{G}\, M_{s}} \le \biggl(\frac{q}{\sqrt{G} \,m}\biggr)^{-1}\,,
\end{equation}
where the equality gives the maximal charge
to mass ratio 
\begin{equation}
\alpha_1= \biggr(\frac{q}{\sqrt{G} \,m}\biggl)^{-1}\,.
\end{equation}
For example, a proton has  a charge to mass ratio
 \mbox{$q_p/\sqrt{G}m_p \sim 10^{18}$}. Using  the Eq. (\ref{limit1})  the proton 
can be added to the  charged star if
\begin{equation}
\label{qmprotons}
\frac{Q_{s}}{\sqrt{G}\,M_{s}} \le 10^{-18}\,.
\end{equation}
Thus, a star with higher charge to mass ratio can not be assembled from protons alone.
On the other hand, if  the infalling particles are  dust particles with a larger charge to mass ratio, the maximum limit for the charge to mass ratio of the  charged star
can be much higher.

An example of this is a self-gravitating ball of FeI nucleus, of mass $M_{ball}$ and charge
$Q_{ball}$. 
As the charge to mass ratio for FeI is $q/\sqrt{G} m \sim 10^{16}$,
the Eq. (\ref{limit1}) for Fe gives
\begin{equation}
\alpha_2=\frac{Q_{ball}}{\sqrt{G}\,M_{ball}} \leq 10^{-16}\,,
\end{equation}
and we see that in this case $\alpha_2 \gg \alpha_1$ (greater than for protons alone, see Eq. (\ref{qmprotons}) above).  

For ``charged dust''  formed by particles
with $q/\sqrt{G} m \sim 1$, a self-gravitating sphere with mass $M_{dust}$
and charge $Q_{dust}$, can be in hydrostatic equilibrium if:
\begin{equation}
\alpha_3=\frac{Q_{dust}}{\sqrt{G}\,M_{dust}} \sim 1\,.
\end{equation}

So : $\alpha_3 \gg \alpha_2 \gg \alpha_1$.
Let us estimate the amount of charge needed to have an extremely charged
$10\, M_{\odot}$ star. In this case  $Q_{star}/\sqrt{G}\, M_{star} =1$, so
\begin{eqnarray}
Q_{star}&&=\sqrt{6.67 \,\,10^{-8}} \times 10 \times 1.9 \,\,10^{33} \nonumber\\
&& =4.9 \times 10^{30} \,\,{\rm statCoulomb}\,.
\end{eqnarray}
Therefore, it is needed to have an excess of 
$$\sim Q_{star}/q_p=10^{40} \,\,\,\,\,\,{\rm charges}\,,$$ 
in the star, where
 $q_p=4.803 \times 10^{-10}\,\,{\rm statCoulomb}$ 
is the charge of one proton.
There is roughly $\sim 10^{58}$ baryons in a $10 M_\odot$ star.
So,  there must be one  charged particle on $\sim10^{18}$ neutral ones
 in order to have  a maximally charged compact object. 
This is compatible
with the limits given above. Therefore, in the extremal case, there are a small charge
imbalance equal to:
\begin{equation}
\delta Q \sim 10^{-18} \,\,\,\, {\rm charges \,\,per\,\, baryon}\,,
\end{equation}
in the star. A tiny  amount of charge from a microscopic point of view, but with a huge total sum (see Ref. \cite{ghezzi2}).

It is possible to make the objection that the nucleus would disintegrate, or suffer nuclear fission, before assembling a charged star. However the energy locally available (in the center of mass of the particles)  from the collapse is not enough to unbind a nucleus (bounded by the strong force). This was calculated for a compact charged neutron star. An exception is a nearly extremal compact object, in this case  is energetically favorable for the nucleus to disintegrate \cite{ghezzi2}. The conclusion is that we can approach to the formation of a ``nearly extreme" object, although it is not 
possible to reach the extremal value, the particles will fission first (see \cite{ghezzi2} for details).

Studying theories with extra dimensions, Mosquera-Cuesta and co-workers \cite{mosquera} found 
that particles located
in the brane can leak out to the bulk space.
It results that electrons can leak out more easily than baryons, 
producing a charge asymmetry that can be very large
in very old stellar systems. This is suitable for type II supernova progenitors
and neutron stars \cite{ghezzi2}. The mechanism produce
a charge imbalance of   \cite{mosquera}:
\begin{equation}
\delta Q \sim 10^{-14}-10^{-16} \,\,\,\,{\rm charges\,\, per\,\, baryon}\,.
\end{equation}
 This is several orders of
magnitude higher than
the needs for producing an extremal star.
Therefore,  charged old stars must be considered and studied as a theoretical possibility.

\section{Equations in co-moving coordinates}
We want to solve the Einstein-Maxwell equations (see Ref. \cite{ghezzi2}):
\begin{equation}
R^{\mu \nu}-\frac{1}{2} g^{\mu \nu} R=\frac{8 \pi G}{c^4} T^{\mu \nu}\,,
\end{equation}
where $R^{\mu \nu}$ is the Ricci tensor, $g^{\mu \nu}$ is the metric tensor,
$R$ is the scalar curvature, and $T^{\mu \nu}$ is the energy momentum-tensor, composed by 
two parts:
\begin{equation}
T^{\mu \nu}=T_{F}^{\mu \nu}+T_{M}^{\mu \nu}\,
\end{equation}
the perfect fluid energy-momentum tensor $T_{F}^{\mu \nu}$ and the Maxwell energy-momentum
tensor $T_{M}^{\mu \nu}$.
The electromagnetic energy-momentum tensor is given by:
\begin{equation}
T_M^{\mu \nu}=\frac{1}{4 \pi} \biggl(F^{\mu \alpha} F^\nu_{\,\,\,\alpha}-\frac{1}{4} g^{\mu \nu} F_{\alpha \beta} 
F^{\alpha \beta} \biggr)\,,
\end{equation}
here $F^{\mu \nu}$ is the Maxwell electromagnetic tensor,
which can be written in terms of a potential $A^{\mu}$:
\begin{equation}
\label{vecpot}
F_{\mu \nu}=A_{\mu;\nu}-A_{\nu;\mu}\,.
\end{equation}
In the last equation and in the rest of the paper, we use semi-colon to denote covariant
derivative.
The perfect fluid energy-momentum tensor is:
\begin{equation}
T_{F}^{\mu \nu}=(P+\delta \,c^2) \,u^\mu u^\nu - P \,g^{\mu \nu}\,,
\end{equation}
where $\delta\,c^2$ is the density of mass-energy and $P$ is the pressure, given in ${\rm [dyn/cm^2]}$,
and $u^{\nu}$ is the $4$-velocity of the observers co-moving with the fluid  \cite{landau}, \cite{weinberg}, \cite{ghezzi2}.
The electromagnetic tensor must satisfy the Maxwell equations \cite{bekenstein}, \cite{mashhoon}, \cite{ghezzi2}:
\begin{equation}
F^{\mu \nu}_{\,\,\,\,\,;\nu}=4 \pi j^{\mu}\,.
\end{equation}
From Eq. (\ref{vecpot}) and  using the Bianchi identities, we have:
\begin{equation}
F_{[\alpha \beta; \gamma]}=0\,.
\end{equation}

We will use the common split of the energy:
\begin{equation}
\delta = \rho\, (1+\epsilon/c^2)\,,
\end{equation}
where $\rho$ is the rest mass  density, and $\rho\,\epsilon$ is the internal energy density \cite{ghezzi2}, \cite{shapiro}.

The fluid must satisfy the energy-momentum conservation equation:
\begin{equation}
T^{\mu \nu}_{\,\,\,\,\,\,;\nu}=0\,,
\end{equation}
and the baryon conservation equation:
\begin{equation}
(\rho \,u^{\nu})_{;\nu}=0\,.
\end{equation}

We will write the Einstein-Maxwell equations
 in coordinates and gauge choice 
appropriated for its numeric implementation
\cite{ghezzi2}.
Considering a spherical star, the line-element in co-moving coordinates is given by
\begin{equation}
ds^2=a(t,\mu)^2 c^2 dt^2-b(t,\mu)^2 d\mu^2-R(t,\mu)^2 d\Omega^2\,. 
\end{equation} 
The 4-velocity of an observer  co-moving with the fluid   is
$u^\nu=[a^{-1},0,0,0]\,,$
and satisfies $u^\nu u_\nu=c^2$, where $c$ is the speed of light.
 The coordinate $\mu$ was gauged to be the rest mass enclosed
by co-moving observers standing on spherical layers of the star. Each 
layer  has its own constant value of $\mu$. In particular, observers at the surface 
of the star will be designated with  a coordinate $\mu=\mu_s$.

 We define\footnote{
This is the $u^1$ component of the co-moving observer's 4-velocity 
in Schwarzschild coordinates, see  Ref. \cite{ghezzi2} for details.}
$$u=R_{,t}/a\,.$$
So, the equation of motion,  obtained from the Einstein-Maxwell equations
\cite{ghezzi2} is
\begin{eqnarray}
\label{motion}
u_{,t}=-a \,\biggl[4\,\pi\, R^2\, \frac{\Gamma}{w}\, \biggl(P_{,\mu} - 
\frac{Q\,Q_{,\mu}}{4\, \pi\, R^4}\biggr)+\frac{G\, m}{R^2}+ \nonumber \\ 
\frac{4\,\pi\, G}{c^2}\,P\, R-\frac{G\, Q^2}{c^2\, R^3} \biggr]\,,
\end{eqnarray}  
where $w=1+\epsilon/c^2+P/\rho c^2$ is the relativistic specific enthalpy, 
$\rho$ is the rest mass density, 
$\epsilon$ is the internal energy per unit mass,
 $P$ is the pressure,  $G$ is the gravitational constant,  $Q(\mu)$ is the total charge integrated from
the origin of coordinates up to the spherical layer with coordinate $\mu$, and $m = m(t,\mu)$ is the
total mass-energy defined below. 
We use the notation $f_{,x}=\partial f/\partial x$, with $x = \{\mu, t\}$,
throughout the paper. 

We note that the Eq. (\ref{motion}) resembles a Newtonian equation of motion plus relativistic corrections.
In the Eq. (\ref{motion}) the last two terms are pure relativistic, and the full equation is equivalent to
its Newtonian counterpart when $c \rightarrow \infty$
\footnote{In the hydrostatic limit we can take $u_{,t}=0$ in the Eq. (\ref{motion})  to obtain 
 the Tolman-Oppenheimer-Volkov
equation for charged stars \cite{ghezzi2}.}.
The factor $\Gamma$ is  a generalization of the special-relativistic $\gamma$ factor for
 general
relativity and is given by
\begin{equation}
\label{gamma}
\Gamma^2=1+\frac{u^2}{c^2}-\frac{2 \,m\, G}{R\, c^2} + \frac{G\, Q^2}{c^4\, R^2}\,.
\end{equation} 
This factor also verifies the equality
$\Gamma=R_{,\mu}/b$  (see Refs. \cite{ghezzi2}, \cite{may}, \cite{bekenstein}).

The equation for the metric component $a=g_{t t}^{1/2}/c$ is obtained from 
the energy-momentum conservation equation $T_{1;\nu}^{\nu}=0$, 
and is given by \cite{ghezzi2} 
\begin{equation}
\label{eqaw}
\frac{(a w)_{,\mu}}{a w}=\frac{1}{w c^2} \biggl[\epsilon_{,\mu} + 
P \,\biggl(\frac{1}{\rho}\biggr)_{,\mu}
+\frac{Q \,Q_{,\mu}}{4 \,\pi\, R^4 \rho} \biggr]\,.
\end{equation}
The other independent component $T_{0;\nu}^{\nu}=0$ gives
\begin{equation}
\label{kineticenergy}
\epsilon_{,t}=-P\,\biggl(\frac{1}{\rho}\biggr)_{,t}\,,
\end{equation}
which is identical to the non-relativistic adiabatic energy conservation
equation, and constitutes the first law of thermodynamics. 
We observe that the \mbox{Eq. (\ref{kineticenergy})} do not contain any electromagnetic
term. This is due to the symmetry of the problem, although an electromagnetic term
arise in the total energy (see the Eq. \ref{totalmass} below).

From the conservation of the number of baryons, 
we obtain the equation for the metric
component $b=g_{\mu \mu}^{1/2}$  \cite{ghezzi2},
\begin{equation}
b(t,\mu)=\frac{1}{4\, \pi\, R^2 \,\rho} \,.
\end{equation}
The Lagrangian coordinate $\mu$ can be chosen to be \cite{ghezzi2}
\begin{equation}
\label{masscoord}
\mu=4\, \pi\,\int_0^{R'} \rho \,R^2 dR/\Gamma\,,
\end{equation}
this is the total rest mass enclosed by a sphere 
of circumference $2\, \pi\, R'$. Thus, the collapsing ball
is divided in layers of constant rest mass $\mu$
and each co-moving observer is at rest in each of this layers
\footnote{The definition of the Lagrangian mass coordinate 
is independent of the presence of
electromagnetic fields \cite{ghezzi2}.}.

The equations for the total mass and the mass conservation are, respectively \cite{ghezzi2}:
\begin{equation}
\label{totalmass}
m(t,\mu)=
4\,\pi\, \int^\mu_0\rho\,\biggl(1+\frac{\epsilon}{c^2}\biggr)\, R^2\, R_{,\mu}\, d\mu\,\, +
\end{equation}
$$\frac{1}{c^2}\int^\mu_0 \frac{Q\, Q_{,\mu}}{R}\, d\mu\,;$$

\begin{equation}
(\rho R^2)_{,t}/\rho R^2=-a \,u_{,\mu}/R_{,\mu}\,.
\end{equation}

The charge 4-current $j^\nu$ is the product of a scalar electric charge density $\rho_{ch}$ 
times $u^\nu$ ;
$j^\nu=\rho_{ch} \,u^\nu$.
It can be shown that the charge and the rest mass are  conserved  
in shells co-moving with the fluid \cite{bekenstein}, \cite{ghezzi2}. 
So, the charge increment
between layers can be written as being proportional to $d\mu$,
\begin{equation}
\label{difcharge}
dQ=4\, \pi\, \rho_{ch}\, R^2\, dR/\Gamma\,,
\end{equation}
where 
\mbox{$\rho_{ch} \propto \rho$}.

We observe that  Eq.  (\ref{difcharge}) is a non-linear equation, since $\Gamma$ 
 depends on the integral value of the charge $Q(\mu)$ (see Eq. \ref{gamma}), 
so the set of equations above must be solved self-consistently.

The equations (\ref{motion})-(\ref{difcharge}) 
above are solved numerically with the following
boundary conditions:
\begin{subequations}
\begin{eqnarray}
&&P=0, \,\,\,{\rm at}\,\,\mu=\mu_{s},\,\,\,\,\forall \,t \,,\label{BCp}\\ 
&&a=1,\,\,\,\,{\rm at}\,\,\mu=\mu_{s},\,\,\,\,\forall \,t \,,\label{BCa}\\ 
&&u=0, \,\,\,\,{\rm at}\,\,t=t_0,\,\,\, \forall \,\mu \,,\label{Bmu}\\ 
&&R=0,\,\,\,{\rm at}\,\,\,\mu=0, \,\,\,\forall \,t\,, \label{BCr}
\end{eqnarray}
\end{subequations}
where $\mu_s$ is the mass coordinate at the surface of the star.
The boundary condition (\ref{BCp}) can be derived from the matching condition between the 
interior and the exterior solution   \cite{ghezzi2}. 
 Equation (\ref{BCa}) express our coordinate freedom to choose the time  synchronized
with a co-moving observer moving with the surface of the star.  Eq. (\ref{Bmu}) means
that the ball is initially at rest (at the initial time $t_0$),
and Eq. (\ref{BCr}) is the impenetrability condition at the origin of coordinates (this
condition can be stated equivalently as $u=0$ at $\mu=0$).
In the next section we will present the initial conditions.

In this work we choose
\begin{equation}
\label{densityofcharge}
\rho_{ch}=constant\,\times\,\rho\,,
\end{equation}
 for simplicity and 
without loss of generality.

\section{Numerical techniques and code setup}
The equations given in the section above are written in a form closely
paralleling the equations of May and White  \cite{ghezzi}, \cite{ghezzi2},  for the non-charged case  \cite{may}, 
\cite{may2}, \cite{voropinov}.

We built the numerical code \mbox{collapse05v2} to integrate the Eqs. (\ref{motion})-(\ref{difcharge}).

The Eq. (\ref{eqaw}) is  re-written  as \cite{ghezzi2}:
\begin{eqnarray}
{a w}&=&{a_0 w_0}\,{\rm exp}\,\biggl[\,\int_0^{\mu_s}\,\biggl( d\epsilon + P \,d \biggl(\frac{1}{\rho} \biggr)+ \nonumber\\
&& \frac{Q \,d Q}{4\, \pi\, r^4\, \rho} 
\biggr) /w\,c^2\biggr]\,. 
\end{eqnarray}
 This equation is integrated numerically  from the surface of the star toward the center,
 using the boundary condition \mbox{$a_s\,w_s=1$}, where $a_s$ and $w_s$ are  
given at the surface of the star (with coordinate $\mu_s$),
note that we already choose $a_s=1$ (see Eq. \ref{BCa}).

The \mbox{collapse05v2} uses a 
leapfrog method plus a predictor
 corrector step, and iterates with a Crank-Nicholson algorithm.
The method is second order accurate in time and space. We use a numerical viscosity
to resolve the strong shock waves formed \cite{may2}.

\subsection{Initial conditions and numerical caveats}
It is assumed a charge density proportional to the rest mass density through the star: 
$\rho_{ch}/\sqrt{G} \rho \sim constant$. From Eqs. (\ref{masscoord}) and (\ref{difcharge})
we see that $Q(\mu)/\sqrt{G}\mu=constant$, $\forall \,(t,\mu)$.

We assumed a  polytropic equation of state, \mbox{$P=k \rho^\gamma$}, where $\gamma=5/3$ is the
 adiabatic coefficient \footnote{For $\gamma=constant$ shocks see \cite{chorin}, pag 128.}. 
The  star has initially a uniform
mass density, $\rho=10^7\,{\rm g\,\,cm}^{-3}$ and a uniform distribution of electric charge 
density, in all the models studied.
The initial specific internal energy $\epsilon$ is varied in the different simulations.
The initial setup imitates a massive star of $M=21.035 M_{\odot}$ that had exhausted its nuclear
fuel. Its fate depends on the internal energy and charge of the initial configuration. This is
 explored with our simulations.
The initial radius of the  star  is 
 $R_0=10^9$ cm, in all the simulations. 

Also we consider that the electric field 
is always much lower than the
Schwinger field limit for
pair creation ($\sim 10^{16}$ Volt/cm, in {\it vacuo} and in flat space-time \cite{Schwinger}). 
For example, in the case of extremal charge $Q/\sqrt{G}M=1$,  the total charge will be
$Q\sim10^{31}$ stat-Coulomb. In this case, the maximum electric field strength is
$\sim10^{11}$ stat-Volt/cm or $\sim 3 \times 10^{15}$ Volt/cm.  
However, during the collapse the field can
strengthen and take higher values, so quantum effects would be taken into account
in a more realistic approach.

We are not considering magnetic fields in the present simulations
\footnote{A warning must be given: a magnetic field 
is induced in the collapsing charged sphere  if perturbed slightly,
and the spherical
symmetry of the problem is broken. 
However, the strength of the induced magnetic field  can be neglected in a first approximation
to the full problem, as long as $|v|^2/c^2 \ll 1$, with $v$ the non-radial component of 
the 3-velocity of the fluid. 
In the present case, $v=0$.}. 

The simulations converge to the same solution when the resolution or the viscosity
parameter are changed.
The  parameters of the simulation were chosen for
 accuracy and efficiency when running on a single processor. 
We performed the  simulations using an Intel Pentium IV 2.6 Ghz processor (32 bits) and compiled
 the code with  \mbox{Lahey/Fujitsu} v6.1 \cite{lahey} for 32 bits; also we used an AMD Athlon FX
 2.6 Ghz processor (64 bits) and compiled with \mbox{Absoft} \cite{absoft} for 64 bits, 
running under Linux
 operating system. The code also runs under the Windows operating system and compile 
 with the  Fortran 90 Power Station. We didn't find essential differences between the different
runs with the exception of the speed of the simulation, which is faster with the AMD processor.
In addition  we set:
a) the number of integration points: 1000; 
b) viscosity parameter: $3.3$; 
c) initial time-step: $10^{-5}\,$sec.
We also performed tests simulations with 300 
and 1500 points to be sure of the physical results, and tested the code varying the viscosity
over a wide range of 
parameters. In the present simulations we tried to minimized the  effects of the viscosity while 
keeping the ability of the code to resolve shock fronts. 

The percentage of numerical error  in  the Hamiltonian constraint, expressed by the conservation 
of the total mass-\mbox{energy $m(t,\mu_s)$},  is given by:
 $$E_H=100\,\times\,|m-m'|/m\,,$$ 
where \mbox{$m'=m'(\mu_s)$} is the exact value of the total mass energy (constant in time)
and \mbox{$m=m(t,\mu_s)$} is the numerical value. The momentum constraint is ``build in" the code and
exactly conserved.
In the present simulations:
$E_H< 0.1\,\%$ when the AH forms, using $1000$ points, and it is $E_H< 0.009\,\%$ with
$1500$ points, with no essential qualitative differences between the solutions. This values of
 $E_H$ are for the worst case of a large binding energy and with very strong shocks, so it can be 
considered as the superior bound of the error. 

After crossing the AH the error grows to: $\sim 0.4 \,\%$ (using $1000$ points), and   
the simulation is stopped. Thus in every simulation we set a maximum  error tolerance 
of $E_H \sim 0.4\,\%$.  We checked that  using a
higher number of points  the errors are reduced. 


We emphasize that the numerical errors (mainly $E_H$, not the roundoff errors) and the intrinsic difficulty on the simulation is the formation of the black hole region, because near it several  numerical operators blows up.

\subsection{The apparent horizon}
For each layer of matter with Lagrangian coordinate $\mu$ 
it is possible to define an  internal and  an external horizon,
\begin{equation}
r_{\pm}(t,\mu)=\frac{G m(t,\mu)}{c^2} \pm \frac{G}{c^2} \sqrt{m^2(t,\mu)-Q^2(\mu)/G}\,,
\end{equation}
which can be interpreted as a generalization of the Schwarzschild radius for charged stars.
During the simulation  an apparent horizon forms
when some layer of matter  cross its external horizon, i.e. when: $r(t,\mu) \le r_{+}(t,\mu)$ \cite{ghezzi2}.
In addition to the formation of the
 apparent horizon, in some cases  a coordinate
singularity develops, i.e.: $g_{tt}(t,\mu') \rightarrow 0$  and $g_{\mu \mu}(t,\mu') \rightarrow +\infty$ (for some
coordinate value $\mu'\ne 0$). Note that at the coordinate $\mu=0$ is located the physical timelike 
singularity of the vacuum  Reissner-Nordstr\"om space-time.
The simulation  must be ended before any observer $\mu$ can meet 
its internal horizon $r_{-}(\mu)$. Hence, we can not
say if the star collapses directly to a singularity or if it pass through a wormhole
 to another asymptotically flat universe.
In this paper, we leave this question open.

The formation of an apparent horizon (AH) 
is a sufficient condition (although not necessary)
 for the formation of an event horizon and a black hole region \cite{ghezzi2}.
Moreover, the interior solution matches with the exterior 
Reissner-Nordstr\"om spacetime
\cite{ghezzi2}.
In this sense, we can say that the outcome of the simulations in which an AH
forms is the formation of a ``Reissner-Nordstr\"om black hole''.
Of course, if not all the matter collapse to the
AH the result at the moment the simulation is stopped
will be an electric black hole plus a surrounding gas with a vacuum 
Reissner-Nordstr\"om  exterior spacetime.

We observe that there must exist other formulations of the problem in which it is
possible to get closer to the singularity while avoiding the AH. 
We will explore this subject in future works.

\section{Discussions of the results}

We performed a set of simulations on the
collapse of a $21\,M_{\odot}$ star varying the initial conditions. 
We found that for certain cases  
a shell-like structure of 
higher  mass density and charge is formed
in the most external part of the star.  
This shell enlarge with time until it reaches the center of the star.

The Table I summarize the results of the simulations performed. 
In each model studied are  di\-ffe\-rent amounts of total electric charge;
 initial internal energy or binding energy. The fate of each model 
depends on this conditions.

 The formation of the shell is sensible to the internal energy, or the binding energy, of the star (see Table I). 
When stronger is the bound of the star 
higher is the wall of the shell.

\begin{figure}[h]
\begin{center}
\epsfig{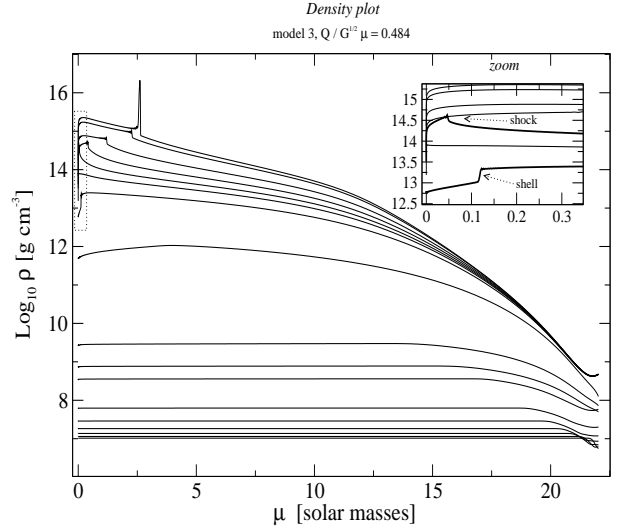} 
\end{center}
\caption{Snapshots of the mass density {\it versus} the
mass coordinate $\mu$, for the model 3 of the Table I, the shell forms mildly in this case, although the shock is
very strong.}
\label{fig:rho3048}
\end{figure}

\begin{figure}[h]
\begin{center}
\epsfig{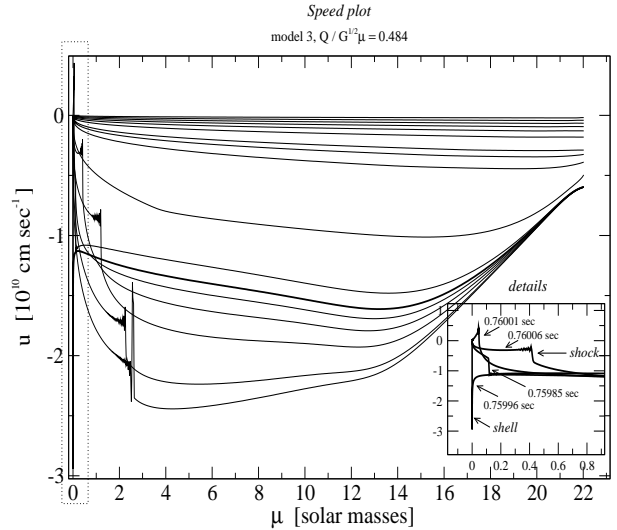} 
\end{center}
\caption{Snapshots of the speed {\it versus} the Lagrangian mass, the inset shows the details of 
the shell impinging on the center and
the shock starting its propagation. The speed is positive during a short lapse of time.}
\label{fig:speed3048}
\end{figure}

\begin{figure}[h]
\begin{center}
\epsfig{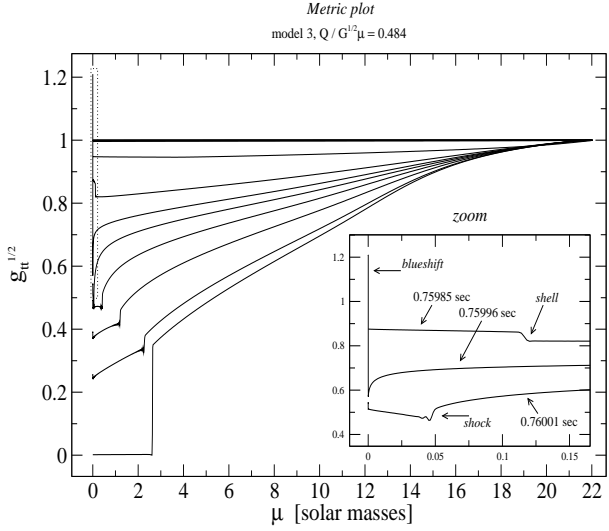} 
\end{center}
\caption{Snapshots of the coefficient $g_{tt}^{1/2}$, it is distinguished a sharp discontinuity at the shock and shell positions. 
The inset shows a zoom of the dotted box, the coefficient $g_{tt}^{1/2}$ is greater than
one at the instant of time the shock revert the sign of the velocity, indicating a blueshift respect 
to the infalling matter.
The shock and the shell are indicated in this figure.}
\label{fig:a3048}
\end{figure}

In the Fig. (\ref{fig:rho3048}), we show the profiles of the mass density {\it versus} the Lagrangian mass $\mu$ for the model $3$ of the Table I, in this
case the shell forms only mildly. However, there is a strong shock wave formed. The Fig. (\ref{fig:speed3048})
shows the velocity profiles, on which it is seen a strong shock propagating outwards. 
We observe that although the shock  is propagating outwards
the sign of the fluid velocity is negative everywhere indicating that the star is always collapsing.
There is an exception in which the shock velocity acquires a positive sign, although it lasts a very brief period of time. This is shown on the 
 inset of the Fig. (\ref{fig:speed3048}): the shell is reaching the coordinate origin, and 
an instant later it impinges the center. After that a shock wave is formed and
 start  to propagate outwards. During a brief lapse of time the velocity is positive, although the shock is not
strong enough to stop the total collapse.
The Figure (\ref{fig:a3048}) contains the snapshots of the metric component $g_{tt}^{1/2}$: it can be seen 
that concordant with 
 the shell impinging  the center,   $g_{tt}$ acquires
a value greater than one, indicating a blueshift with respect to the infalling matter. The inset of the 
Fig. (\ref{fig:a3048}) shows
the details of the jumps produced on the value of $g_{tt}^{1/2}$ at the shell and at the shock positions, as well as
the mentioned blueshift.

\begin{figure}[h]
\begin{center}
\epsfig{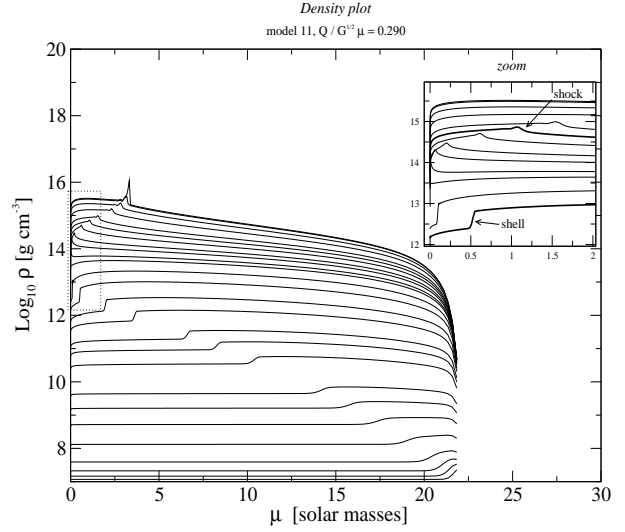} 
\end{center}
\caption{Snapshots of the mass density {\it versus} $\mu$, for the model 11 of the Table I.
In this case there is an important imploding shell. The dotted box is zoomed on the inset,
which shows the shell 
impinging at the center and the shock forming and propagating outwards.}
\label{fig:rho3029}
\end{figure}

\begin{figure}[h]
\begin{center}
\epsfig{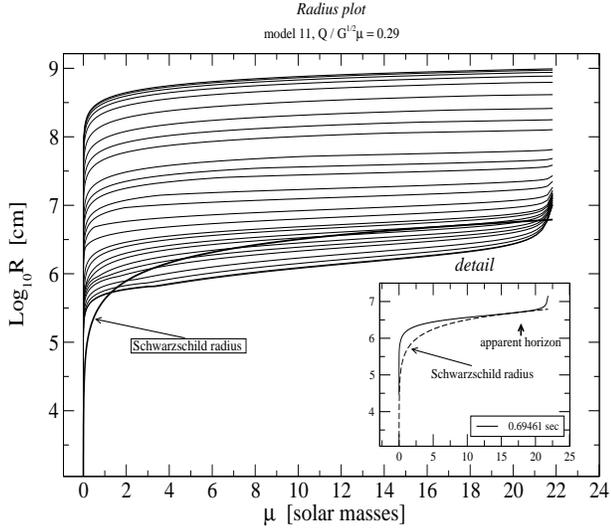} 
\end{center}
\caption{
The radius (Log$_{10}$Radius) {\it versus} the Lagrangian mass, the bold line
shows the Schwarzschild radius at the instant of the apparent horizon formation,
the inset box shows the detail of an observer crossing his/her own Schwarzschild radius
(an arrow is pointing the observer) and signaling
the formation of an apparent horizon. It can be appreciated that the simulation continues 
far beyond the formation of the apparent horizon.}
\label{fig:r3029}
\end{figure}

 Figure (\ref{fig:rho3029}) contains the snapshots of the mass density for the model $11$ of the Table I. In this case
there is an strong shell formed. On this figure we can see that the shell propagates inwards, to lower values
of the Lagrangian coordinate $\mu$, until it impinges the center. At this moment, and place, it is formed 
an strong shock that starts its outward propagation.  The inset box of the Figure (\ref{fig:rho3029}) is a zoom
of the dotted box, showing the details of the collapsing shell and the shock formed. 
The Fig. (\ref{fig:r3029}) shows the verification that the collapse produce a black hole. In spherical symmetry the 
apparent horizon is formed simply when an observer cross his/her own Schwarzschild radius. The inset box
of the Fig. (\ref{fig:r3029}) shows the radius profile at the moment that the curve touch 
the Schwarzschild radius profile at a point. That point indicates an observer crossing the Schwarzchild radius
and is the \mbox{\it locus} of the apparent horizon. After then the apparent horizon will evolve 
approaching the event horizon at infinite coordinate time or after complete collapse (not calculated).

\begin{figure}[h]
\begin{center}
\epsfig{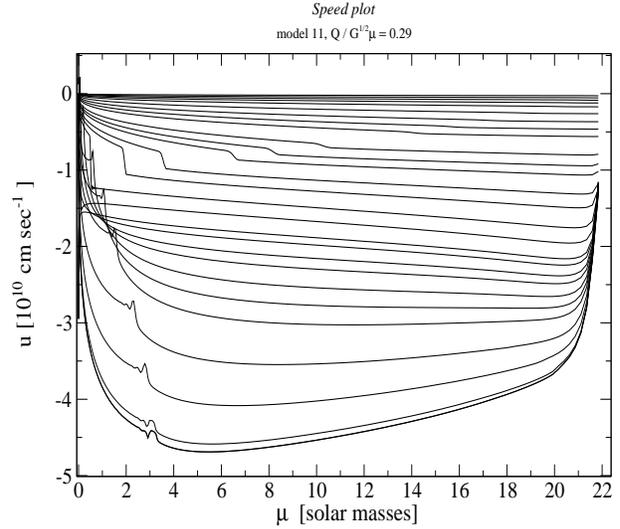} 
\end{center}
\caption{Snapshots of the speed {\it versus} $\mu$ for model 11, it can be observed a jump
in the velocity profile at the shell position dislocating through the center of the star.
See Fig. (\ref{fig:speed3029zoom}).}
\label{fig:speed3029}
\end{figure}

\begin{figure}[h]
\begin{center}
\epsfig{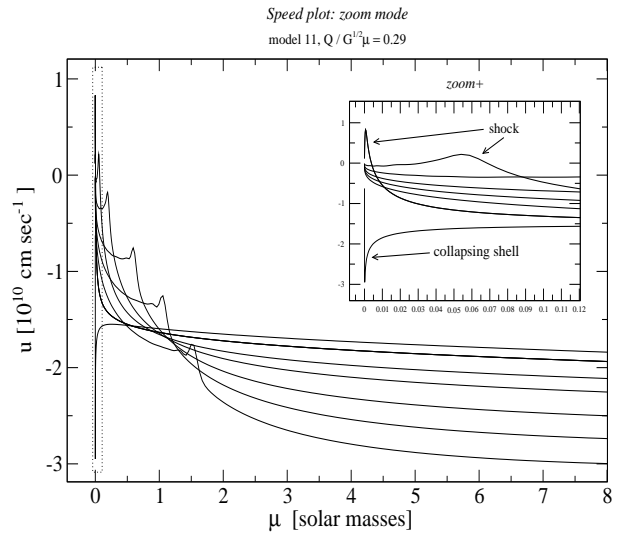} 
\end{center}
\caption{ Detail of the speed's profile showing the moment of time the shell impinges at the center 
of the star, later the shock is 
formed and starts its outward propagation, the inset shows a zoom of the dotted box. 
See Fig. (\ref{fig:speed3029}).}
\label{fig:speed3029zoom}
\end{figure}

\begin{figure}[h]
\epsfig{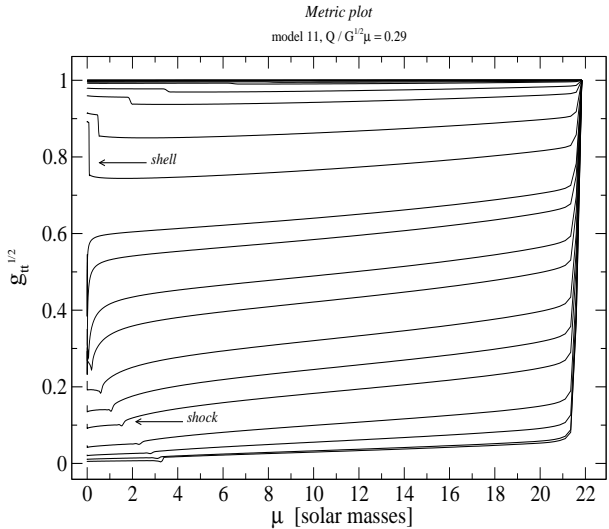} 
\caption{ Temporal snapshots for $\sqrt{g_{tt}}$ {\it versus} $\mu$ for model 11 and its late collapse
to zero. We must observe that the apparent horizon forms before $g_{tt} \rightarrow 0$, which 
indicates the convergence towards the event horizon.
 }
\label{fig:a3029}
\end{figure}

 Figure (\ref{fig:speed3029}), shows the velocity profiles for model 11, there is an strong-shell 
and strong-shock in this case.
However, the velocity has a 
negative sign and the matter collapse directly to a black hole.
 Figure (\ref{fig:speed3029zoom}) shows the speed profile at the moment of time in which the shell impinges  the center, and
the formation and propagation of the shock wave. The inset box shows a close up of the details. There is a short
lapse of time in which the shock has positive velocity.
The Fig. (\ref{fig:a3029}) shows the metric coefficient $g_{tt}^{1/2}$ going to zero. In this case $g_{tt}$
 never becomes greater than one, and the inner matter is always redshifted respect to observers standing at the
surface or outside the star.

\begin{figure}[h]
\epsfig{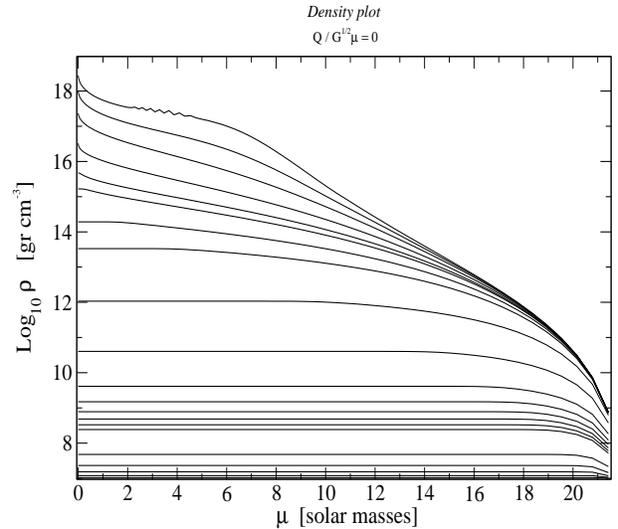} 
\caption{ Snapshots of the mass density {\it versus} $\mu$ for model 13, 
an uncharged  star.}
\label{fig:rho00}
\end{figure}

\begin{figure}[h]
\epsfig{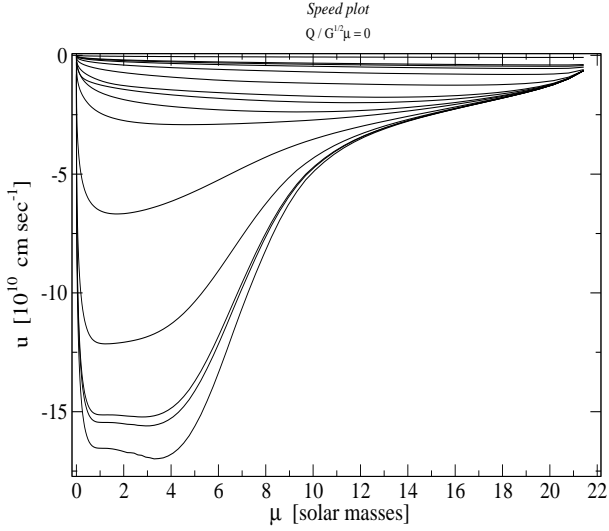} 
\caption{ Snapshots of the speed profile, where $u$ has the meaning of the
rate of radius change per unit proper (comovel) time change.
}
\label{fig:speed00}
\end{figure}

\begin{figure}[h]
\epsfig{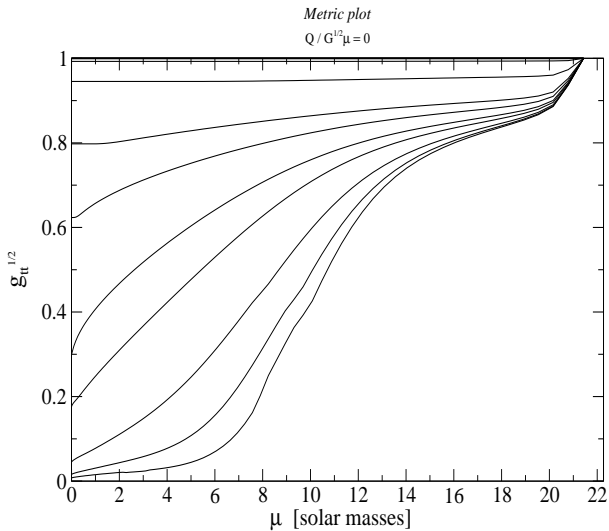} 
\caption{ Temporal snapshots for $\sqrt{g_{tt}}$ {\it versus} $\mu$ for model 13.}
\label{fig:a00}
\end{figure}

\begin{figure}[h]
\epsfig{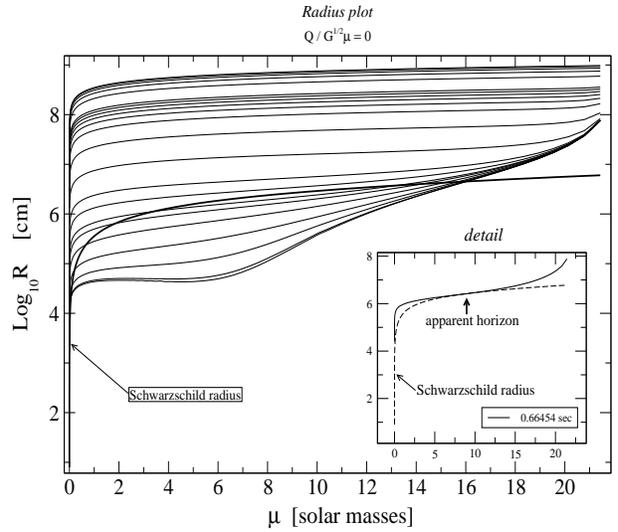} 
\caption{ The radius {\it versus} $\mu$: the bold line shows the Schwarzschild radius at the instant of time the apparent
horizon is formed, the inset shows the detail of an observer, pointed with an arrow, crossing his/her
own Schwarzschild radius, indicating the formation of an apparent horizon at this coordinate 
and at the time indicated.}
\label{fig:r00}
\end{figure}

 Figure (\ref{fig:rho00}) shows the density snapshots for the collapse of an uncharged star (model $13$). It is possible to compare this
simulation with the former simulations, we see that for the charged collapse the density profiles are flatter all
the way through the black hole formation.  Figure (\ref{fig:speed00}) shows the velocity profiles 
(compare it with the Figs. \ref{fig:speed3048} and \ref{fig:speed3029}).  
The Figure (\ref{fig:a00}) shows the collapse of $g_{tt}^{1/2}$ 
(compare it with the Figs. \ref{fig:a3048} and \ref{fig:a3029}).  Figure (\ref{fig:r00}) shows the profiles
of the metric function $R$, the bold line is the Schwarzschild radius at the moment the apparent horizon forms. The inset box shows the detail of the radius profile touching the Schwarzschild profile at one point, signaling the formation of an apparent horizon at the point indicated with the vertical arrow, at the time indicated on the legend box. The simulation is continued beyond that event, as seen in this figure.

\begin{figure}[h]
\epsfig{width=8cm,height=7cm, file=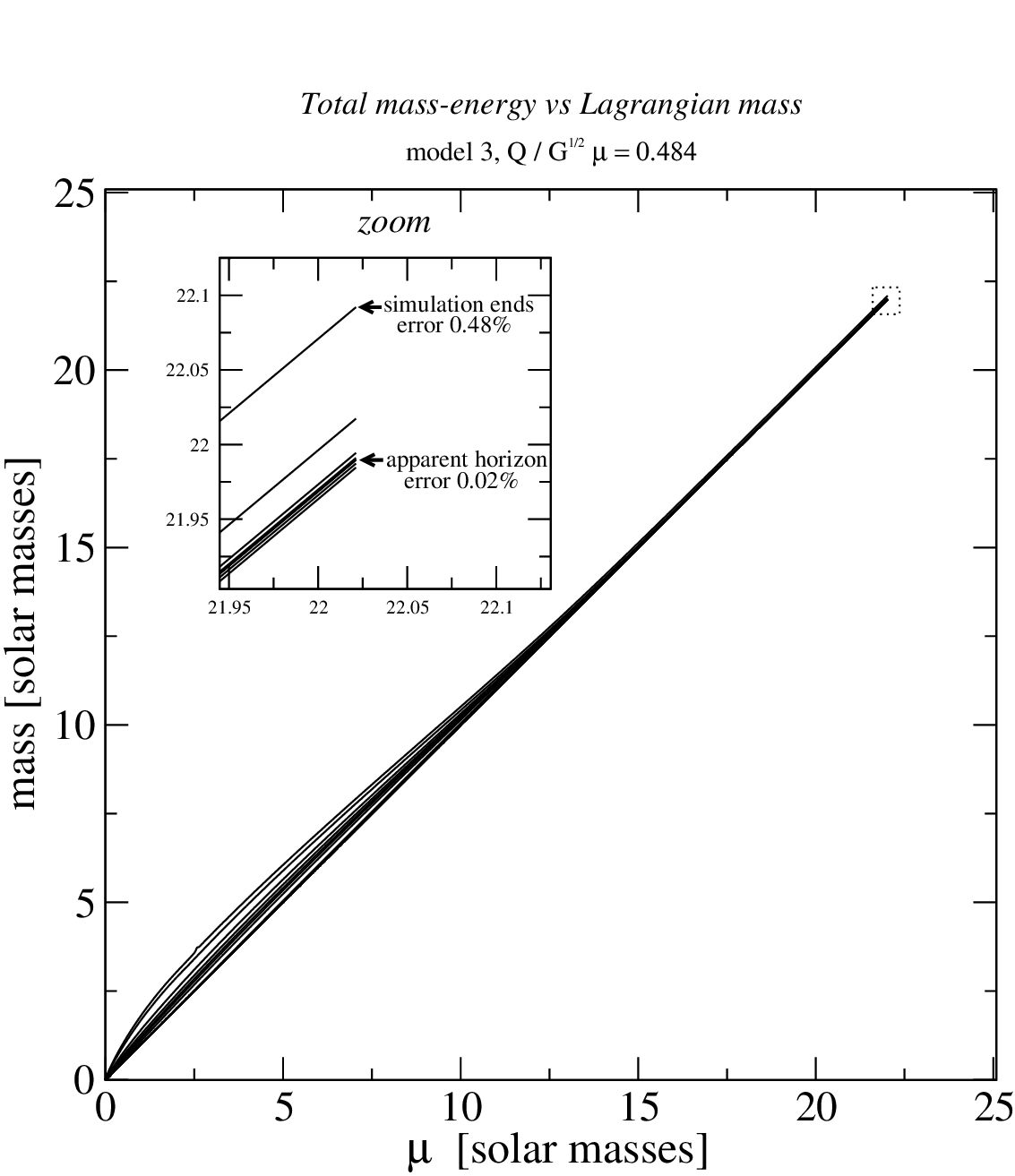} 
\caption{ Snapshots for the mass-energy  {\it versus} $\mu$ for model 3. The numerical error in the simulation of the models can be measured
by the failure of the curves to converge at the same point, this is indicated on the zoom
boxes. After the apparent horizon formation the error is $0.02\,\%$ for model 3 and model 13. The 
same error is found for model 13 using a lower number of radial points (see text for details). 
Compare with Fig. (\ref{fig:massa00}).}
\label{fig:massa3048}
\end{figure}

\begin{figure}[h]
\epsfig{width=8cm,height=7cm, file=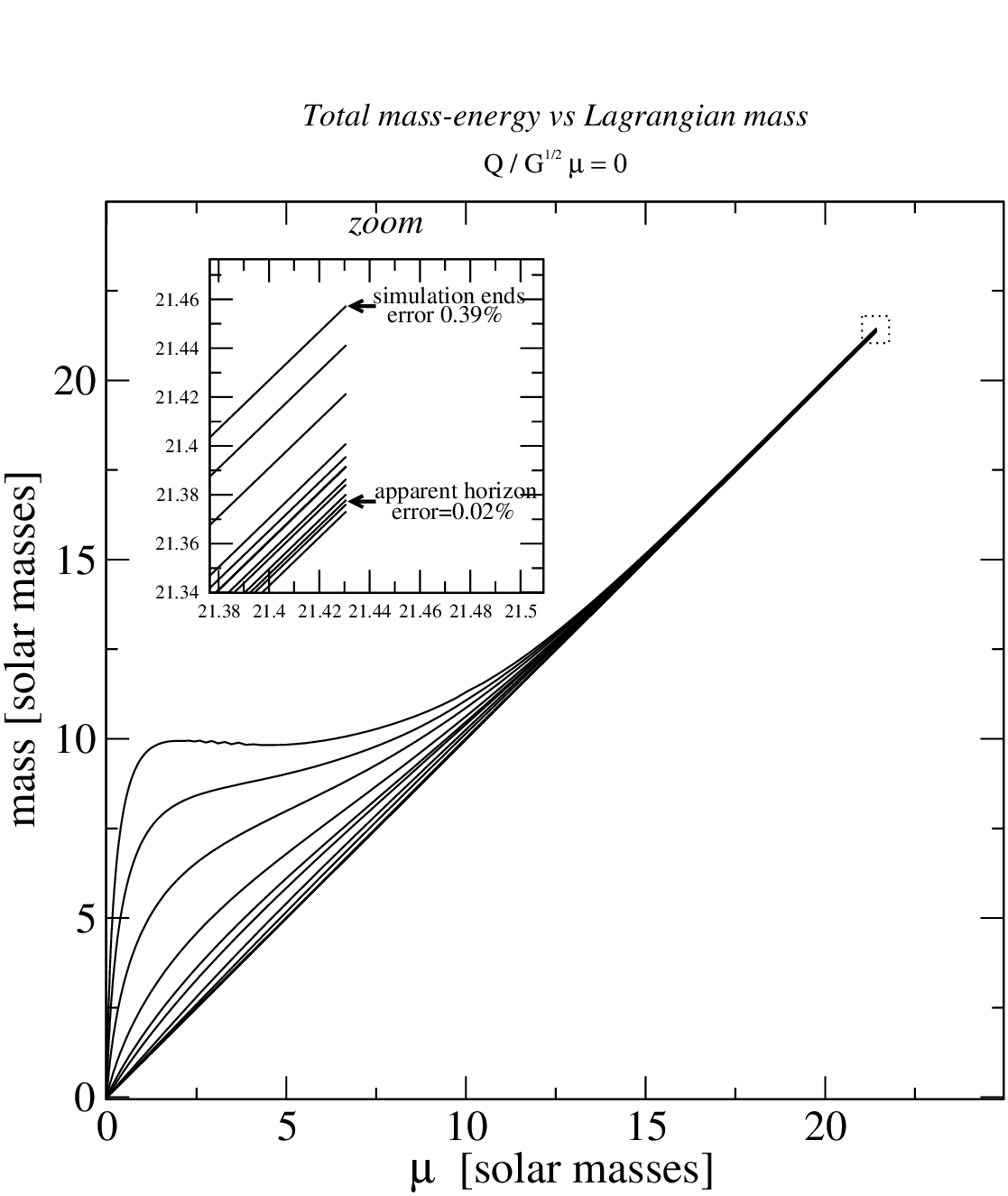} 
\caption{ Snapshots for the mass-energy  {\it versus} $\mu$ for model 13. 
The numerical error in the simulation of the models can be measured
by the failure of the curves to converge at the same point, this is indicated on the zoom
boxes. After the apparent horizon formation the error is $0.02\,\%$ for model 3 and model 13. The 
same error is found for model 13 using a lower number of radial points (see text for details).
Compare with Fig. (\ref{fig:massa3048}).}
\label{fig:massa00}
\end{figure}

 It is possible to compare the profile of the mass-energy function for the charged collapse with very strong shocks (model 3) Fig. (\ref{fig:massa3048}), with the case of uncharged collapse (model 13) Fig. (\ref{fig:massa00}). The inset box of the Fig. (\ref{fig:massa3048}) shows a zoom of the right end of the curves which indicates the level of energy conservation in the simulation. All curves must reach the same point for perfect energy conservation. At the moment of the apparent horizon formation the mass-energy is conserved with a precision of 
$99.98\,\%$, which is an excellent value taking into account that this is the worst case.
After the apparent horizon formation, the good conserving property is lost as in any relativistic code, and the simulation is ended with a $0.48\,\%$ error. We emphasize that this is the physical quantity we use
to check the convergence properties of our code, since the momentum constraint is conserved exactly by the algorithm.
As was said before the simulations were performed using $1000$ points, and the conservation can be better using a larger number of points.
The Fig. (\ref{fig:massa00}) represents the energy conservation for the case of uncharged collapse. The precision is roughly the same as in the former case, although using only $100$ points. We checked that using $1000$ points the errors are lowered by a factor of $10$ in this case, i.e.: $99.998\,\%$ precision at the apparent horizon formation. The reason for  better conservation of the energy constraint is that there are not very strong shocks like in the former simulations.

The explanation of the shell formation can be grasp from the equation of motion  (Eq. \ref{motion}). 
This effect can also happen in Newtonian physics, but its evolution in the strong field regime is 
highly non-linear and far from obvious.
It seems difficult to make a complete analytic description of its evolution.
However, starting with a ball of constant rest mass density  and charge, it is easy to see that a shell must be formed. 
The relativistic term $G\, Q^2/c^2\, R^3$, in the Eq. (\ref{motion}), produce a repulsion of the matter on the initial
homogeneous sphere, which in turn produce a charge gradient ${Q Q_{,\mu}}/{4\, \pi\, R^4}$.

The term
${Q Q_{,\mu}}/{4\pi R^4}$ in the equation of motion  can be important  in supporting 
the weight of the star  (see Eq. \ref{motion}). 
 Near the surface of the star,  the pressure gradient term  $P_{,\mu}$ 
cancels the charge gradient term ${Q\,Q_{,\mu}}/{4\, \pi\, R^4}$  at a certain point we call $\mu_0$. 
Since at this point the two terms cancels out, all the matter outside the layer of coordinate
$\mu_0$ is almost in free fall, producing an accumulation of matter and giving rise to the shell, while
at coordinates with $\mu < \mu_0$ the matter is supported by a positive net gradient. 
We took care to check that this is not a numerical artifact, changing the boundary conditions at the surface, the amount of charge, and the initial conditions.

Hence, it results that the point $\mu_0$ is roughly the {\it locus in quo}
the term  $P_{,\mu}(\mu_0)$ and the term ${Q Q_{,\mu}}/{4\pi R^4} (\mu_0)$ sum zero, then  
the gas is compressed and a shell of matter forms.

The non-linearity in the strong field regime comes from the fact that $\Gamma$ is a function
of the charge and the mass and multiplies the two gradient terms, enhancing the effect (see Eq. \ref{motion}).

From all the numerical experiments we performed, we observed that the shell formation effect arise more clearly when the initial
density profile is flat. For the case of charged neutron star collapse the effect is 
negligible  \cite{ghezzi2}.

 The point $\mu_0$
shift to lower values with time, and this brings the shell to enlarge until $\mu_0=0$, when it reaches
the origin and rebounds forming the shock wave mentioned above. 
It is observed that the shock do not form in the uncharged case, using the
same set of initial conditions.
The density contrast between the shell and its interior is higher for greater total charge (or 
equivalently higher internal electric field
and zero total charge). This is  because 
the Coulomb repulsion inside the shell will be higher, and this gives a higher 
compression of the matter at the shell.
The Coulomb repulsion term together with the pure relativistic 
term $a {G Q^2}/{c^2 R^3}$ in the Eq. (\ref{motion}), are responsible for the 
positive mass density gradient near the 
coordinate origin.
The charged collapse is slightly delayed respect to the uncharged one, 
i.e.:$\sim 0.75$ sec for model 1 respect to
the uncharged case that lasts $\sim 0.66$ sec.
At the end of the simulation, we observe that the density profile looks globally flatter 
and the maximum density value is lower than in the uncharged case.

The case 14 in Table I is special because it represents stars with a charge
greater or equal than the extremal value. We verified that in this cases the star 
expands ``forever", i.e.:  we follow its expansion until the density 
gives a machine underflow. In any of these
cases we expect a re-collapse of the matter: the
binding energy per nucleon tends to zero when the charge tends to the mass (see \cite{ghezzi2}), 
and is negative for a charge greater than mass.

\subsection{The maximum mass limit}
For charged stars there must exist a maximum mass limit like for uncharged compact stars. 
For white dwarfs the limit is known as Chandrasekhar mass limit, while for neutron stars is
the Oppenheimer-Volkoff limit, etc.
In this section, we calculate the mass  limit for Newtonian charged compact stars.
We assume, for simplicity, an equation of state dominated by 
 electrons \footnote{See Ref. \cite{shapiro} for the uncharged case.},
\begin{equation}
P/\rho=Y_e k T/m_B+K_\gamma Y_e^\gamma \rho^{\gamma-1}\,.
\end{equation}
For a spherical configuration, the Newtonian hydrostatic 
 equilibrium gives: 
\begin{equation}
P_c/\rho_c=GM/R-Q^2/(4 \pi R^4 \rho)\,.
\end{equation}
Simplifying the two equations,  for $T=0$ and $\gamma=4/3$ we get 
$(G M^2-Q^2)/M^{4/3}=K Y_e^{4/3}$.
From this equations, we obtain the limiting mass  $M_{max}$. Assuming the charge is a 
fraction $\alpha$ of the mass, 
i.e.: $Q=\alpha \sqrt{G} M$, 
\begin{equation}
M_{max}=\biggl( \frac{K}{G (1-\alpha^2)}\biggr)^{3/2} Y_e^2 \sim 5.83 \frac{Y_e^2}{(1-\alpha^2)} M_\odot\,,
\end{equation}
which reduces to the known Chandrasekhar mass formula when $\alpha=0$ (no charge), modulo some geometric 
factor depending on the density profile.

For the extreme case $\alpha=1$, the configuration disperse to infinity. We checked that this also
happens in the relativistic case (see model 14 of Table I).

\subsection{Implications for core collapse supernovae}
The quantity of energy deposition behind the formed shock wave is critical to produce 
a successful supernova explosion, like in the proposed mechanisms for core collapse supernovae
explosions
\cite{colgate}, \cite{Bethe}, \cite{wilson}. 
In the cases we studied the shock wave is not fast enough to reach the surface 
of the star before the apparent horizon forms ahead of it, and the shock  ends trapped by the black hole region.
However, when including neutrino transport in our simulations the shell formation could have
a dramatic effect.

This provide a new mechanism for a successful 
core collapse supernova explosion. 
Although, the calculations performed in the present work are far from a complete supernova
calculation, it is simple to show that in the charged collapse there are a more efficient  energy
deposition to revive the supernova shock wave.

As the density profile is flatter than in the uncharged
case, the neutrino-sphere and gain radius for neutrino deposition  must be different. 
In order to show this, we consider the change in the radius $R_\nu$, of the neutrino sphere, which is defined by $\tau(R_\nu)=2/3$, where
the optical depth of the uncharged matter for the neutrinos is \cite{Bethe}
\begin{equation} 
\tau_{}(R)=1.5 \times 10^{11}\, (N^2/6A)\, \epsilon_{\nu}^2\, R^{-2}.
\end{equation} 
According to the numerical results presented, 
we make the approximation $\rho \sim constant$, for the charged case, in order to obtain 
an analytic expression for the optical 
depth for neutrinos. In this case, the 
 optical depth of the charged matter is
\begin{equation}
\tau_{ch}(R)=10^{-8}\, (N^2/6A)\, \epsilon_{\nu}^2\, \rho_{12}\, (R_s-R). 
\end{equation} 
It can be seen that $\tau_{ch}\gg\tau$.
Assuming typical values \cite{Bethe}: $\epsilon_{\nu}=10$ Mev  and
 $\rho \sim 2 \times 10^{10}\,{\rm g\, cm}^{-3}$,
in the uncharged case the position of the neutrino-sphere is at: $R_\nu \sim 110$ km, and in the charged 
case \footnote{ From the
simulations, we obtain the radius  $R_s = 954$ km, at which 
$\rho=2\times 10^{10}\,{\rm g\, cm}^{-3}$, 
 for the particular case $Q/\sqrt{G}M=0.48$.} it is:
$R_\nu \sim 845$ km. This implies a more efficient neutrino trapping during the charged collapse,
and hence a more efficient energy deposition.

\begin{table*}
\caption{\label{tab:table4} Simulations of the collapse of a massive star
varying the charge to mass ratio and the binding energy.}

\begin{ruledtabular}
\begin{tabular}{rcccccccc}

Simulation & charge to mass ratio & $\epsilon$\footnotemark[1]  & 
  Binding energy\footnotemark[2]  & collapse ? & shell ?&
 \\
& $Q / \sqrt{G}\, m$& [${\rm ergs}/c^2$] &        \\ 
\hline 
\hspace{0.10cm}
1 &  $0.484$ & $9.61\times 10^{-7}$  &    $-2.994 \times 10^{-2}$ & yes & yes &\\
2 &  $0.484$ & $1.92\times 10^{-6}$  &    $-2.992 \times 10^{-2}$ & yes & no  &\\
3 &  $0.484$ & $2.88\times 10^{-6}$  &    $-2.990 \times 10^{-2}$ & yes & no  &\\
4 &  $0.484$ & $1.44\times 10^{-3}$  &    $+2.498 \times 10^{-4}$ & no  & no  &\\
5 &  $0.484$ & $1.35\times 10^{-3}$  &    $-1.764 \times 10^{-3}$ & no  & no  &\\
6 &  $0.145$ & $1.15\times 10^{-6}$  &    $-3.830 \times 10^{-2}$ & yes & no  &\\
7 &  $0.145$ & $9.61\times 10^{-8}$  &    $-3.833 \times 10^{-2}$ & yes & no  &\\
8 &  $0.145$ & $9.61\times 10^{-9}$  &    $-3.833 \times 10^{-2}$ & yes & yes &\\
9 &  $0.290$ & $9.61\times 10^{-7}$  &    $-3.583 \times 10^{-2}$ & yes & no  &\\
10 & $0.290$ & $9.61\times 10^{-8}$  &    $-3.584\times 10^{-2}$  & yes & yes &\\
11 & $0.290$ & $9.61\times 10^{-10}$ &    $-3.585\times 10^{-2}$  & yes & yes &\\
12 & $0.850$ & $9.61\times 10^{-7}$  &    $-1.073 \times 10^{-2}$ & yes & yes &\\
13 &  $0$    & $9.61\times 10^{-7}$  &    $-3.740 \times 10^{-2}$ & yes & no  &\\
14 &$\ge 1.0$& $\dots$               &    $\dots$                 &  no & no  &\\

\end{tabular}
\end{ruledtabular}

\footnotetext[1]{The energy density is initially distributed uniformly on the star in all the simulations.
The equation of state is  $P=k\, \rho^\gamma$ with $\gamma=5/3$. The mass of the star in every simulation is
$M=21.035 \,\,M_{\odot}$.}

\footnotetext[2]{The binding energy of the star:  $(m-\mu) \,c^2$, in units of 
${\bf M}_{\odot}\,c^2$ ergs.}

\end{table*}

\section{Final Remarks}
We have performed relativistic numeric simulations for the collapse of spherical symmetric stars
 with a polytropic equation of state and 
possessing a uniform distribution of electric charge.
We have not studied in this paper how the matter acquires a high internal electric field before or 
during the collapse process.

 In  the present model we studied the  essential features of 
the collapse of stars with a total electric charge.
The simulations
are an approximation to the more realistic problem of the temporal evolution of
an star with an strong internal electric field and total charge zero. This  model is under consideration by the authors.

In the cases were the stars formed an apparent horizon  it is unavoidable the formation of a Reissner-Nordstr\"om black hole. This is guaranteed by  the singularity theorems and by the Birkhoff theorem.
  
For low values of  the charge to  mass ratio
$Q /\sqrt{G} M \ll 1$,  no difference  with the collapse of an uncharged star was found.
The value of the total charge that prevent the collapse of the star with any initial condition 
is given by a charge to mass ratio $Q/\sqrt{G}M \ge 1$.  
That means that stars with $Q/\sqrt{G}M \ge 1$ spreads 
 and do not collapse to form black holes nor stars in hydrostatic equilibrium.
It is observed a dramatically different physical behavior
when $Q/\sqrt{G} M > 0.1$.  In this case, the collapsing matter forms a 
shell-like structure, or bubble,  surrounding
an interior region of lower density and charge.  The effect is due to the competence between 
the Coulomb electrostatic repulsion and the attraction of gravity. 
This effect must occur in non-relativistic physics as well.  
The relativistic case is more interesting due to the
 non-linearity of the Einstein equations. 
The effect is more important when lower is the internal energy of the star (see Table I).
In some of the experiments we observe a blueshift produced at the bouncing shock wave (see Fig. \ref{fig:a3048}), because the strong shock wave acquires a positive velocity over a small lapse of time. 

For all the cases studied, the density profile is globally flatter than in 
the uncharged collapse.

The optical depth for neutrinos 
is  much higher in the case of the charged collapse, hence, the neutrino trapping must be more efficient.  
In conclusion, if the internal electric field increase during the stellar collapse the formation of the 
shell must be taken seriously in supernova simulations.

In addition, we obtained the mass limit formula for Newtonian charged stars, 
which clearly precludes the formation of naked singularities.  
The mass limit can not be naively extrapolated for the relativistic case, because
charge and pressure regeneration effects can change the maximum charge to mass ratio.
However, in the present work, we checked 
that it is not possible to form an star from matter with a charge to mass ratio 
greater or equal than one in agreement with our mass formula and with the cosmic censorship conjecture.

A more complete simulation
including neutrino transport, and other quantum effects, 
is being considered by the authors.

\section*{Acknowledgments}
The authors acknowledge the Brazilian agency FAPESP. 
This work was benefited from useful discussions with Dr. Samuel R.  Oliveira and Dr. Eduardo Gueron. We acknowledge Gian Machado de Castro for a critical reading of the manuscript.  
The simulations were performed at the Parallel Computer Lab.,  Department of Applied Mathematics,  UNICAMP 
(Campinas State University).


\end{document}